# THE TIMING SYNCHRONIZATION SYSTEM AT JEFFERSON LAB[*]

M. Keesee, R. Dickson, R. Flood, Jefferson Lab,
12000 Jefferson Ave, Newport News, VA 23606, USA
V. Lebedev, Fermi National Accelerator Laboratory
P.O. Box 500, Batavia, IL 60510

Abstract

This paper presents the requirements and design of a Timing Synchronization System (TSS) for the Continuous Electron Beam Accelerator Facility (CEBAF) control system at Thomas Jefferson National Accelerator Facility. A clock module has been designed which resides in a VME crate. The clock module can be a communications master or a slave depending on its configuration, which is software and jumper selectable. As a master, the clock module sends out messages in response to an external synchronization signal over a serial fiber optic line. As a slave, it receives the messages and interrupts an associated computer in its VME crate. The application that motivated the development of the TSS, the Accelerator 30 Hz Measurement System, will be described. Operational experience with the Timing Synchronization System will also be discussed.

## 1 INTRODUCTION

The CEBAF accelerator is a five pass CW recirculator, which can reach an energy of 6.067 GeV. It consists of a 67 MeV injector, two superconducting 600 MeV linacs, and 9 arcs of magnets which connect the linacs for beam recirculation. See figure 1.

In order to improve machine reproducibility and reduce beam tune time, the 30 Hz Measurement System was developed. By means of a small set of correctors or RF cavities, a beam perturbation can be induced at a frequency of 30 Hz. The effect of this perturbation is apparent at any Beam Position Monitor (BPM) point in the machine.

The BPM software was enhanced to detect beam differential displacements resulting from such a 30 Hz perturbation of lateral beam position or energy. For beam tuning, AC line synchronized 60 Hz pulsed beam is used. The BPM system performs data acquisition at 60 Hz synchronized with this pulse. The choice of 30 Hz as the perturbation frequency enables synchronization of every other beam pulse to the positive or negative crest of the perturbation signal. It thus becomes important for the BPM systems to be able to differentiate positive or negative crested data, i.e. the BPMs must maintain synchronization with the 30 Hz perturbation polarity.

Prior to the development of the hardware-based TSS, the implementation of the 30 Hz measurements used a software based synchronization system. This was an accelerator control system network broadcast to inform the BPMs of the polarity of the 30 Hz signal. This approach resulted in a system sensitivity to network delays that occasionally caused the BPM system to label the data with the wrong polarity. It therefore became apparent that a different approach was needed.

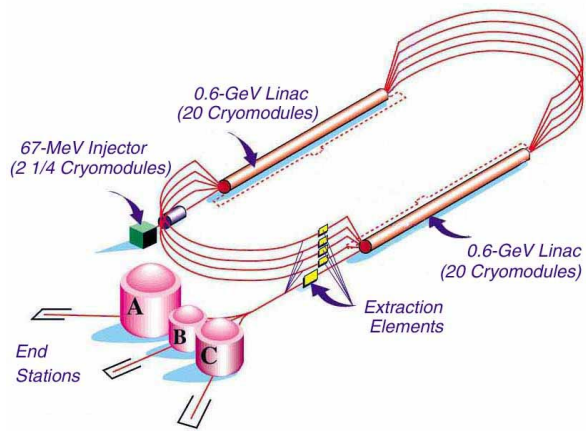

Figure 1. Machine Configuration

## 2 SYSTEM OVERVIEW

The hardware-based TSS uses one master clock module and multiple slave clock modules [1]. These 3U sized cards share the VME bus with the control system's front end Input-Output Controllers (IOCs). There are IOCs at many locations around the accelerator. The communication flow, from master to slave boards, occurs over a single fiber optic line originating from the master. Each card has a fiber optic receiver and repeating transmitter. The last card in the chain is connected back to the master. This enables the master to perform a continuity test of the entire link. The frequency of the message passing is AC line synchronized 60 Hz.

---

[*] This work was supported by the U.S. DOE contract No. DE-AC05-84-ER40150

The message from the master to the slave timing modules contains the status and polarity of the 30 Hz perturbation signal sent to the correctors or cavities. The BPM system reads this message and performs the appropriate measurement calculations.

Since the sole usage of the fiber link is for timing synchronization and there is only one source of messages, this system guarantees synchronization to a time approximately equal to the maximum round trip time for a message. Worst-case delay from received input to re-transmitted output per module is less than 1.5usec. In a complete implementation, with up to 100 IOCs participating, the total round trip time will be 150usec. This eliminates the possibility of receiving incorrect information regarding the beam perturbation polarity.

## 3 TIMING MODULE

The purpose of the master timing module is two-fold. First, it produces a TTL compatible square wave at 30 Hz. Phase locked to this signal, an HP3314A function generator modulates corrector magnets or RF cavities and provides the desired beam perturbation. Second, at each transition of this square wave, the master transmits on the fiber link, a message indicating the polarity of this square wave. The slave timing module makes this information available to each BPM system. The BPM system, which gathers data at 60 Hz, tags the data with the slave supplied polarity.

The accelerator has its own 60 Hz AC line synchronized global timing reference called the Beam Sync. Numerous data acquisition systems throughout the accelerator, including the BPM system, use this signal for data acquisition. The master 30 Hz clock module also uses the Beam Synch signal in order to maintain phase synchronization with the accelerator.

In summary, the Beam Sync synchronizes, at 60 Hz, the BPM system and the master timing card. The master timing module outputs a Beam Sync phase locked 30 Hz TTL square wave to drive the function generator and provides phase information (i.e. polarity) to the BPMs, via the fiber link of this square wave.

Upon reception of a message from the master module, the slave module will issue a VME interrupt to the IOC. The interrupt request level is jumper-selectable, as is the master or slave status of the module. The interrupt vector number is software programmable and is part of the IOC's system initialization.

The clock module is a 24 bit address, 16-bit data slave with D08 vectored-interrupt capability. The VME bus interface, twelve 8-bit registers, and the serial message encoding/decoding are implemented in a 144-pin Altera ACEX EP1K50 Field Programmable Gate Array (FPGA) operating at 20Mhz. The device is re-programmable via a front-panel JTAG interface [2].

Several diagnostics have been developed to ensure proper operation of this system. Software in each IOC will detect the loss of interrupts from the timing module if they fail to occur at a 60 Hz rate. This will cause the BPM software to issue an alarm invalidating 30 Hz reported data. A system expert can further diagnose the system by means of self-test software. A quick determination of fiber link status may be obtained by examining the LED on the front panel of the timing module. This LED will light if the link carrier is detected and will blink if messages are being received.

## 4 30 HZ MEASUREMENT SYSTEM

The 30 Hz Measurement System was developed in order to improve machine reproducibility and reduce beam tune time. This system makes it possible to track and correct machine optics by measuring the differential response generated by perturbations of the beam.

Two types of modulation devices are used for two different types of measurement. Beam energy perturbations are performed by changing the accelerating gradient of the superconducting cavities. Transverse beam modulation is performed by adjusting horizontal or vertical air core dipole corrector magnets. Both types of modulation can be applied at either at the end of the injector at an energy of 67MeV, or at the beginning of the first arc at an energy of 667 MeV, providing optics measurements of the entire machine [3]. An important feature of the system is that it is possible to induce the perturbations and make these measurements not only during accelerator beam tuning, but also during accelerator operations. This is possible because the Fast Feedback system [4] removes these small perturbations prior to beam delivery to the experiments.

The BPM system includes more than 850 BPMs, controlled by two different types of electronics, distributed across a network of 30 IOCs. The Switched Electrode Electronics (SEE) BPM [5] software uses the new TSS to determine the 30 Hz modulation polarity and uses this information in the data acquisition calculations. As mentioned earlier, the 60 Hz Beam synch triggers the data acquisition. The software then calculates the mean value of the beam position and the amplitude of the 30 Hz beam motion.

# 5 PRESENT STATUS AND FUTURE PLANS

To date, timing modules have been installed in nine locations around the accelerator. The system has recently come on-line and at present is working as designed. Future upgrades to the module's FPGA firmware and IOC software will allow non BPM IOCs to benefit from the installation of this module. Eventually, all VME crates with IOCs (approximately 100) will have a clock module, with BPM systems having the highest installation priority.

In the very near future, two additional functions will be implemented within the Timing Synchronization System. These are absolute time of day synchronization and remote system reset capability.

Time of day synchronization will be implemented using a new timing module message type and using network Ethernet based TCP/IP communication. The new message type will be a time hack type message. The approach will be to broadcast, via Ethernet, to all IOCs that a time hack type timing module message is imminent and upon reception of this message, the IOCs should synchronize their time of day clocks to the time specified in this Ethernet broadcast. The IOC's on-board timer will maintain the time reference between these hacks. This system will allow time of day synchronization at sub-millisecond level.

For the future, additional serial-data encoding methods are being investigated, along with the possibility of increasing the module's on-board clock speed, which should reduce the per-board serial data latency to less than 1usec.

Since the timing module is a standard VME card, it has access to the VME bus System Reset line. Asserting this line will cause the associated IOC to reboot and all boards in that crate to reset. A new message type will be created to command a timing module to assert this reset. All timing modules have a jumper-selectable eight-bit address allowing 255 unique addresses (address 0 is for broadcasting messages to which all slaves respond). Reset messages will allow targeting individual VME crates for resets. By bit mask partitioning of the addresses, it is possible to reset a subsystem of IOCS. For example, it will be possible to reset all BPM IOCs with a single command. At present, many IOCs are at inconvenient locations and must be manually reset should they crash. This feature will minimize the impact of such a crash.